\documentclass[prl,twocolumn,superscriptaddress]{revtex4}
\usepackage{graphicx}
\usepackage{graphics}      
\usepackage{bm}    
\usepackage{amsmath}     
\usepackage{amsfonts}
\usepackage{amssymb}
\usepackage{latexsym}  
\begin{document}
\title{FISHER INFORMATION: QUANTUM UNCERTAINTY RELATION}
\newcounter{count}
\author{I.Chakrabarty}
\email{indranilc@indiainfo.com} \affiliation{Heritage Institute of
Technology, Kolkata, India} \affiliation{Bengal Engineering and
Science University, Howrah, India}
\date{\today}
\begin{abstract}
The paper deals with the reformulation of quantum uncertainty
relation involving position and momentum of a particle on the
basis of the Kerridge measure of inaccuracy and the Fisher
information.
\end{abstract}
\maketitle
\section{\bf Introduction}
A basic problem in quantum physics is to find the limits placed on
the joint measurability of non-commuting variables which may be e.
g., the optical phase and the photon numbers in optics, the
position and momentum of a free and bound particle in atomic
physics. The fact that two non-commuting observables A and B
cannot simultaneously obtain sharp eigenvalues represents the
cornerstone of the principle of uncertainty in quantum mechanics
and this can be expressed in different forms, commonly called
uncertainty relation [1]. An uncertainty relation provides an
estimate of the minimum uncertainty expected in the outcome of a
measurement of an observable, given the uncertainty in outcome of
a measurement of another observable. The limits of the
measurability placed on the concrete quantum system are commonly
given by standard and entropic uncertainty relations [2-5]. The
standard uncertainty relation of two non commuting observables
represents the product of their standard deviations whereas the
entropic uncertainty relation is given by the sum of their Shannon
entropies.\\
In the present paper, we shall use the concept of statistical
inference and the Fisher information to reformulate the
uncertainty relation of non-commuting observables. The problem is
to discuss the quantum uncertainty relation involving position and
momentum of a particle on the basis of the Kerridge measure of
inaccuracy (or the Kullback relation information) and the Fisher
information [6-8].
\section{\bf Quantum mechanics and the Fisher information}
A fundamental problem of statistical inference is the problem of
deciding how well a set of outcomes, obtained in independent
measurements, fits to a proposed probability distribution [9] if
the probability is characterized by one or more parameters. This
problem is equivalent to inferring the value of its parameter(s)
from the observed measurement outcomes x. Perhaps the simplest and
the most well-known approach to the studied problem is the theory
of estimation developed by R.A. Fisher [10]. In this approach, it
is assumed that one out of a family of distribution functions
$[P_{\theta}(x),\theta\in R]$ is the true one, the parameter
$\theta$ being unknown. To make inferences about the parameter
$\theta$ one constructs estimators, i.e. the functions
$T(x_1,x_2,...,x_n)$ of the outcomes of n independent repeated
measurements. The value of the function is intended to represent
the best guess for $\theta$ . Several criteria are imposed on
these estimators in order to ensure that the values do in fact
constitute 'good' estimates of the parameter $\theta$. One
criterion is unbiasedness. For example, if
\begin{eqnarray}
\langle T\rangle=\int_{R}T(x_1,x_2,...,x_n)\prod_{i=1}^n
p_{\theta}(x_i)dx_i
\end{eqnarray}
for all $\theta$,that is, if the expectation value of T represents
that value, we call the estimator $T(x_1,x_2,...,x_n)$ to be an
unbiased estimator of $\theta$. Again, if the standard deviation
$\sigma(T)$ is as small as possible, the estimator is called
efficient. The famous Cramer-Rao inequality puts to a bound to the
efficiency of an arbitrary estimator [6]:
\begin{eqnarray}
Var(T)=\sigma^2(T)\geq \frac{(\frac{d\langle
T\rangle}{d\theta})^2}{nI(\theta)}
\end{eqnarray}
where
\begin{eqnarray}
I(\theta)=\int_{R}(\frac{\partial ln
p_{\theta}(x)}{\partial\theta})^2p_{\theta}(x)dx
\end{eqnarray}
is a quantity depending only on the family
$[P_{\theta}(x),\theta\in R]$, known as the Fisher information.
According to Fisher [10], $I(\theta)$ is the amount of information
about the parameter $\theta$ contained in the random variable
$\tilde{x}$ [10],in the case of a single observation (2) reduces
to
\begin{eqnarray}
Var(T)=\sigma^2(T)\geq \frac{(\frac{d\langle
T\rangle}{d\theta})^2}{I(\theta)}
\end{eqnarray}
and, finally, if the estimator T is unbiased, the inequality (4)
becomes
\begin{eqnarray}
Var(T)=\sigma^2(T)\geq \frac{1}{I(\theta)}
\end{eqnarray}
In quantum mechanics, the probability amplitudes, and not the
probability densities, are the fundamental quantities.
Accordingly, we define the Fisher information in quantum mechanics
as follow :\\
Let $[\psi_{\theta}(x),\theta\in R]$ be the family of Schrodinger
wave functions sufficiently well behaved with respect to the
parameter $\theta$. The parameter $\theta$  may be interpreted as
temporal a spatial shift or any other physical parameter.
According to the statistical interpretation of wave function
$p_{\theta}(x)=|\psi_{\theta}(x)|^2$ describes the probability
density of the particle, if $\psi_{\theta}(x)$ is normalized. The
wave function $\psi_{\theta}(x)$ is a probability amplitude
corresponding to $p_{\theta}(x)$ (for any real $\phi$,
$exp(i\phi)P_{\theta}(x)\psi_{\theta}(x)$ is also a probability
amplitude corresponding to $p_{\theta}(x)$ ). The Fisher
information of $[\psi_{\theta}(x),\theta\in R]$ with respect to
the parameter $\theta$, is defined according to (3) as [11,12].
\begin{eqnarray}
I(|\psi_{\theta}(x)\rangle)=\int_{R}(\frac{\partial ln
|\psi_{\theta}(x)|^2}{\partial\theta})^2|\psi_{\theta}(x)|^2dx\nonumber\\=4\int_{R}(\frac{\partial
ln |\psi_{\theta}(x)|}{\partial\theta})^2dx
\end{eqnarray}
provided the integral is finite. This is essentially the Fisher
information of the family of likelihood functions $p_{\theta}(x)
=|\psi_{\theta}(x)|^2$ . In particular, if we assume the
invariance of the wave function under the shift of location
parameter x that is, if $\psi_{\theta}(x)=\psi(x+\theta)$ then (6)
becomes
\begin{eqnarray}
I(|\psi_{\theta}(x)\rangle)=4\int_{R}(\frac{\partial
|\psi(x+\theta)|^2}{\partial\theta})^2dx\nonumber\\=4\int_{R}(\frac{d|\psi(x)|}{dx})^2dx
\end{eqnarray}
which is now independent of the parameter $\theta$ and henceforth
we shall denote it by $I(\psi)$. In the next section we shall
study the deep significance of the Fisher information $I(\psi)$
and the Cramer-Rao inequality in relation to uncertainty relation.
\section{\bf Fisher information: uncertainty relation}
For the sake of simplicity, we consider a one-dimensional system ó
a particle whose quantum state is represented by Schrodinger wave
function $\psi(x)$. Particle's co-ordinate x, in the statistical
interpretation of the wave function $\psi(x)$, is a continuous
variable with probability density
\begin{eqnarray}
P(x)=\psi^{*}(x)\psi(x)=|\psi(x)|^2
\end{eqnarray}
The co-ordinate x and momentum p of the particle, according to the
Heisenberg uncertainty principle, are subject to the uncertainty
relation.
\begin{eqnarray}
(\Delta x)(\Delta p)\geq \frac{\hbar}{2}
\end{eqnarray}
where $(\Delta x)$ and $(\Delta p)$ are the standard deviations of
the position (location) x and momentum p respectively. For
simplicity we assume that the center of the wave packet is at x =
0, that is, $\langle x \rangle=0$ and let $\langle p \rangle=0$.
Then [13,14]
\begin{eqnarray}
(\Delta x)^2=\langle x \rangle^2=\int_{R}x^2|\psi(x)|^2dx\\
(\Delta p)^2=\langle p
\rangle^2=\int_{R}x^2|\frac{\hbar}{i}\frac{d\psi(x)}{dx}|^2dx
\end{eqnarray}
Let us now approach to the uncertainty relation (9) by a route
based on the Fisher information and the Cramer-Rao inequality
developed in the statistical theory of estimation [6,7,10]. Stam
[15] was the first who pointed out the importance of the Fisher
information and the Cramer-Rao inequality in the study of quantum
uncertainty relation. We are going to do this but with a
difference. Our method is based on the Kullback relative
information [7] and the Kerridge measure of inaccuracy in the
choice of correct probability density [8].\\
Any measurement of the position x is always a subject to an error.
Let $\hat{x}=x+\partial x$ be the observed value of the position
x, where $\partial x$ is the inaccuracy in the location parameter
x resulting from the measurement. Then according to Kerridge [8],
the error occurred about the state of the system (particle) in
accepting the probability density $P(x+\partial x)$ in place of
P(x) is given by [7]
\begin{eqnarray}
K(x+\partial x)=\int_R P(x)\ln\frac{P(x)}{P(x+\partial
x)}dx\nonumber\\=\int_R
|\psi(x)|^2\ln\frac{|\psi(x)|^2}{|\psi(x+\partial x)|^2}dx
\end{eqnarray}
The expression (12), known as the Kullback-Leibler discrimination
information or simply the Kullback relative information, gives a
measure of directed divergence between the probability densities
$P(x)$ and $P(x + \partial x)$. For small $\partial x$, expanding
$K(x + \partial x)$ in powers of $\partial x$, we have
\begin{eqnarray}
K(x+\partial x)=\frac{1}{2}I(\psi)(\partial x)^2
\end{eqnarray}
where
\begin{eqnarray}
I(\psi)=\int_R
[\frac{d}{dx}\ln|\psi(x)|^2]^2|\psi(x)|^2dx\nonumber\\=4\int_R
[\frac{d}{dx}|\psi(x)|]^2
\end{eqnarray}
is the Fisher information with respect to the position x. In
general $\psi(x)$ is a complex function, but in the particular
case when it may be a real function [12]
\begin{eqnarray}
I(\psi)=4\int_R [\frac{d}{dx}|\psi(x)|]^2
\end{eqnarray}
Our basic problem is to reduce the error about the state of the
system given by (13) resulting from the measurement. This can be
achieved by the Cramer-Rao inequality [6]
\begin{eqnarray}
I(\psi)(\partial x)^2\geq \frac{(\partial x)^2}{(\Delta x)^2}
\end{eqnarray}
where $(\Delta x)^2=\langle (x-\langle x \rangle)^2\rangle=\langle
x^2 \rangle$ is the mean square deviation of the position of the
particle.Since
\begin{eqnarray}
(\frac{\hbar}{i})^2I(\psi)=4\int_R
|\frac{\hbar}{i}\frac{d\psi(x)}{dx}|^2dx=4\langle p^2\rangle
\end{eqnarray}
we can reduce the inequality (16) to the usual form of the
Heisenberg uncertainty relation :
\begin{eqnarray}
\langle x^2\rangle\langle p^2\rangle\geq \frac{\hbar}{2}
\end{eqnarray}
or
\begin{eqnarray}
(\Delta x)(\Delta p)\geq \frac{\hbar}{2}
\end{eqnarray}
provided we take the inaccuracy $\partial x$ in the location
parameter x to be equal to the standard deviation $(\Delta x)$.
The equality in (18) corresponds to the equality in (16). The
later holds when
\begin{eqnarray}
\frac{d}{dx}\ln|\psi(x)|^2=\alpha(x-\langle x\rangle)=\alpha x
\end{eqnarray}
or
\begin{eqnarray}
|\psi(x)|^2=A\exp(-\lambda x^2)
\end{eqnarray}
Adjusting the constants A and $\lambda$ by the normalization
condition we have the Gaussian wave packet [13,14]
\begin{eqnarray}
\psi(x)=\frac{1}{\sqrt{2\pi(\Delta x)^2}}\exp[-\frac{x^2}{4(\Delta
x )^2}]
\end{eqnarray}
which corresponds to the wave packet having minimum uncertainty
product
\begin{eqnarray}
(\Delta x)(\Delta p)=\frac{\hbar}{2}
\end{eqnarray}
The above approach is different from that of Stam [15] and others
[11,12]. In the present case the Fisher information is not the
starting concept, it results from the Kerridge measure of
inaccuracy in terms of the Kullback relative information. The
uncertainty relation (18) results from the requirement of reducing
the inaccuracy in the measurement process.
\section{\bf Conclusion}
There exists extensive literature on the different forms of the
uncertainty relations in quantum mechanics [16]. The present paper
is an attempt to re interprets the traditional quantum uncertainty
relation in terms of the statistical theory of information and
inference. The basis of the present approach is the Kerrdge's
interpretation of the Kullback relative information (12) as a
measure of inaccuracy. The Kullback relative information
$K(x|x+\partial x)$ given by (13) introducing the Fisher
information $I(\psi$ defines a metric - a statistical distance on
the parametric space. The importance of statistical distance in
the study of uncertainty relations (both thermodynamical and
quantum mechanical) was stressed by Uffink and Van Tith [17].
\section{\bf Acknowledgement}
The Author wishes to thank Prof C.G.Chakraborti,S.N.Bose Professor
of Theoretical Physics,Department of Applied
Mathematics,University of Calcutta and Prof B.S. Choudhuri,
Department of Mathematics, Bengal Engineering and Science
University for the encouragement and inspiration. The author
acknowledges Prof C.G.Chakraborti for being the source of
inspiration  in carrying out research.He also thanks the learned
referee for the valuable comment which helps the better exposition
of the paper.
\section{\bf Reference}
$[1]$ W. Heisenberg: The Physical Principles of Quantum Theory,
New York, Dover Publication, 1949\\
$[2]$ V. Majernik, L. Richterek: Eur. J. Phys.18 (1997) 79 \\
$[3]$ V. Majernik, E. Majernikova: J. Phys. A: Math. Gen.35
(2002)5751\\
$[4]$ V. Majernik, E. Majernikova, S. Shpyrko: Central Eur. J.
Phys.3 (2003) 393\\
$[5]$ H. Massen, J.B.M. Uffink: Phys. Rev.60 (1988) 1103\\
$[6]$ H. Cramer: Mathematical Methods of Statistics, Princeton,
Princeton University Press, 1946\\
$[7]$ S. Kullback: Information Theory and Statistics, New York,
Wiley and Sons,1959 [8]\\
$[8]$ D.F. Kerridge: J. Roy. Stat. Soc. Ser. B23 (1961) 184\\
$[9]$ I. Vajda: Theory of information and statistical decision,
Alfa, Bratislava, 1981\\
$[10]$ R.A. Fisher: Proc. Cambridge Phil. Soc.22 (1925) 700\\
$[11]$ B.R. Friden: Physics from Fisher Information, Cambridge,
Cambridge University Press, 1998\\
$[12]$ S. Luo: J. Phys. A: Math. Gen.35 (2002) 5181 \\
$[13]$ L.L. Schiff: Quantum Mechanics, New York, McGrew Hill Book
Co., 1955\\
$[14]$ J.L. Powell, B. Caseman: Quantum Mechanics, Massachusetts,
Addision Wesley Publication, 1963\\
$[15]$ A.J. Stam: Information and Control 2 (1959) 101 \\
$[16]$ J.B.M. Uffink: Measurements of uncertainty and the
uncertainty principle, PhD. thesis, University of Utrecht, 1990\\
$[17]$ J.B.M. Uffink, J. Van Tith: Foundation of Physics29 (1999)
655
\end{document}